\documentclass[prl,aps,twocolumn,showpacs,floatfix]{revtex4-1}
\usepackage{amssymb}
\usepackage{hyperref}
\usepackage{graphicx}
\usepackage{dcolumn}
\usepackage{bm,amsmath,verbatim}
\usepackage{mathrsfs}
\usepackage{color}

\def\be{\begin{equation}}
\def\ee{\end{equation}}
\def\bea{\begin{eqnarray}}
\def\eea{\end{eqnarray}}
\def\bse{\begin{subequations}}
\def\ese{\end{subequations}}

\def\be{\begin{eqnarray}}
\def\ee{\end{eqnarray}}

\begin{document}

\title{Dirac and Weyl Rings in Three Dimensional Cold Atom Optical Lattices}
\author{Yong Xu}
\author{Chuanwei Zhang}
\email{chuanwei.zhang@utdallas.edu}
\affiliation{Department of Physics, University of Texas at Dallas,
Richardson, Texas 75080, USA}
\begin{abstract}
Recently three dimensional topological quantum materials with gapless energy
spectra have attracted considerable interests in many branches of physics.
Besides the celebrated example, Dirac and Weyl points which possess gapless
point structures in the underlying energy dispersion, the topologically
protected gapless spectrum can also occur along a ring, named Dirac and Weyl
nodal rings. Ultra-cold atomic gases provide an ideal platform for exploring
new topological materials with designed symmetries. However, whether Dirac
and Weyl rings can exist in the single-particle spectrum of cold atoms
remains elusive. Here we propose a realistic model for realizing Dirac and
Weyl rings in the single-particle band dispersion of a cold atom optical
lattice. Our scheme is based on previously experimentally already
implemented Raman coupling setup for realizing spin-orbit coupling. Without
the Zeeman field, the model preserves both pseudo-time-reversal and
inversion symmetries, allowing Dirac rings. The Dirac rings split into Weyl
rings with a Zeeman field that breaks the pseudo-time-reversal symmetry. We
examine the superfluidity of attractive Fermi gases in this model and also
find Dirac and Weyl rings in the quasiparticle spectrum.
\end{abstract}

\maketitle

The topology of band structures plays a crucial role in many important
phenomena~\cite{volovik,GeimRMP,Kane2010RMP,Qi2011RMP} in various physical
fields, ranging from solid-state materials to photonic crystals, and to cold
atomic gases. Remarkably, apart from gapped topological insulators that
exhibit metallic edge states protected by symmetries~\cite%
{Kane2010RMP,Qi2011RMP}, materials with gapless band dispersions could also
possess non-trivial topological properties and protected edge states. A
well-known example of two-dimensional gapless materials is graphene with
Dirac points~\cite{GeimRMP}. In recent years, gapless Dirac and Weyl points
in three dimensions (3D) have been theoretically predicted~\cite%
{Young2012PRL, Wan2011prb,Yuanming2011PRB,Burkov2011PRL,
ZhongFang2011prl,LingLu2013NP,Dai2014,Hasan2015} and experimentally observed
\cite{Hasan2014NC,Borisenko2014PRL,Liu2014Science,Lu2015,Xu2015,Lv2015} in a
variety of Dirac and Weyl semimetals. Besides isolated topological gapless
points, the gaps of energy spectra in 3D could also close along a line,
forming Dirac and Weyl nodal rings in nodal semimetals~\cite%
{Burkov2011PRB,Carter2012PRB,Chen2015NC,
Kim2015PRL,Chen2015arXiv,Rappe2015arXiv,
Cava2015arXiv,Liang2015arXiv,Rui2015arXiv,HasanRing}. The dispersion of the
surface states in these nodal semimetals vanishes, suggesting a potential
type of high temperature superconductivity~\cite{Volovik2011PRB}.

Cold atomic gases provide a clean platform for discovering new topological
quantum materials due to their high controllability for engineering
Hamiltonians with desired symmetries and versatile tools for directly
probing topological states. In this context, recent experimental
achievements for realizing cold atom topological matter, both gapped and
gapless, mainly focus on low dimensional (2D or 1D) systems, including the
realization of topological Haldane model~\cite{Esslinger2014Nature} for
quantum anomalous quantum Hall effects, the observation of Zak phase~\cite%
{Bloch2013NP} and topological charge pumping in optical superlattices~\cite%
{Bloch2015arXiv,Takahashi2015arXiv,Spielman2015}, the realization of Dirac
cones in optical lattices~\cite{Esslinger2012Nature} and spin-orbit-coupled
gases~\cite{Jing2015arXiv}. In 3D, however, various topological gapless
structures such as Weyl points, structured Weyl points, Weyl rings, and
structured Weyl rings have only been theoretically predicted in the
quasiparticle spectrum of superfluids~\cite%
{Gong2011prl,Melo2011PRL,Das2013PRB,Sumanta2013PRA,Yong2014PRL,Dong2014arXiv, LiuBo2014arXiv,Yong2015}%
. One exception is the Weyl points that have been proposed in the
single-particle spectrum of moving lattices~\cite{Tena2015RPL,Law2015}.
However, a realistic scheme for realizing Dirac and Weyl rings in the
single-particle spectrum of cold atoms is still lacking and their
corresponding superfluid properties have been unexplored.

In this paper, we propose an experimental scheme for engineering a
Hamiltonian that hosts gapless Dirac or Weyl rings in its single-particle
spectrum of cold atoms. The scheme is based on the experimentally already
implemented Raman coupling setup for spin-orbit coupling \cite%
{Lin2011Nature,Jing2012PRL,Zwierlen2012PRL,PanJian2012PRL,Qu2013PRA,Spilman2013PRL}%
, and therefore should be experimentally feasible. Our main results are:

(i) We construct a new spin-dependent Hamiltonian in the continuous space
and derive its tight-binding form. Such Hamiltonian has not been discussed
previously in solid-state materials \cite%
{Burkov2011PRB,Carter2012PRB,Chen2015NC,
Kim2015PRL,Chen2015arXiv,Rappe2015arXiv,
Cava2015arXiv,Liang2015arXiv,Rui2015arXiv}. The Hamiltonian preserves both
pseudo-time-reversal and inversion symmetries without Zeeman fields,
allowing the existence of Dirac rings. The pseudo-time-reversal symmetry is
broken with a Zeeman field and a Dirac ring splits into two Weyl rings. The
parameter regions as well as the topological characterization (e.g.,
topological invariance, surface states) for these topological gapless rings
are obtained.

(ii) We investigate the superfluidity of attractive Fermi gases in this
Hamiltonian and find two distinct superfluid phases (SF1 and SF2). The
transition between them is the first order. Interestingly, Dirac and Weyl
rings also exist in the quasiparticle spectra in certain region of the
superfluid.

(iii) The spin-dependent Hamiltonian can be realized using an experimental
setup based on previous Raman coupling scheme for spin-orbit coupling \cite%
{Lin2011Nature,Jing2012PRL,Zwierlen2012PRL,PanJian2012PRL,Qu2013PRA,Spilman2013PRL}%
. Specifically, two pairs of Raman laser beams are used to couple two
hyperfine spin states of atoms for generating a specific spin-dependent
optical lattice, which is essential for the creation of these topological
nodal rings.


\textit{Model Hamiltonian}: We start from a spin-dependent Hamiltonian in
the continuous space that can support the existence of nodal rings%
\begin{equation}
H=\frac{\mathbf{p}^{2}}{2m}-\sum_{\nu =x,y,z}V_{\nu }\cos ^{2}(k_{L\nu
}r_{\nu })+h_{z}\sigma _{z}-V_{SO}\sigma _{y},  \label{WeylH}
\end{equation}%
where $\mathbf{p}=-i\hbar \nabla $ is the momentum operator, $m$ is the mass
of atoms, $V_{\nu }$ and $a_{\nu }=\pi /k_{L\nu }$ are, respectively, the
strength and period of a periodic lattice along the $\nu $ direction, $h_{z}$
is the Zeeman field, $\sigma _{\nu }$ are Pauli matrices for spins, and $%
V_{SO}=\Omega _{SO}\sin (k_{Lx}r_{x})\cos (k_{Ly}r_{y})\cos (k_{Lz}r_{z})$
corresponds to a spin-dependent optical lattice. For simplicity, we explore
the physics of this Hamiltonian in the tight-binding model (see the
supplementary information for its derivation and comparison with the
continuous model) that can be written as
\begin{equation}
H_{TB}=H_{h}+H_{Z}+H_{SO},  \label{HTB}
\end{equation}%
where $H_{h}=-\sum_{j}\sum_{\sigma }\sum_{\nu }(t_{\nu }\hat{c}_{j,\sigma
}^{\dagger }\hat{c}_{j_{\nu }+1,\sigma }+t_{N\nu }\hat{c}_{j,\sigma
}^{\dagger }\hat{c}_{j_{\nu }+2,\sigma }+h.c.)$ includes the
nearest-neighbor (NN) and next nearest-neighbor (NNN) hopping with the
tunneling amplitudes $t_{\nu }$ and $t_{N\nu }$, respectively, $%
H_{Z}=h_{z}\sum_{j}(\hat{c}_{j,\uparrow }^{\dagger }\hat{c}_{j,\uparrow }-%
\hat{c}_{j,\downarrow }^{\dagger }\hat{c}_{j,\downarrow })$ is the Zeeman
field term, and $H_{SO}=it_{\text{SO}}\sum_{j}(-1)^{j_{x}+j_{y}+j_{z}}(\hat{c%
}_{j,\uparrow }^{\dagger }\hat{c}_{j_{x}+1,\downarrow }-\hat{c}_{j,\uparrow
}^{\dagger }\hat{c}_{j_{x}-1,\downarrow })+h.c.$ is the position-dependent
spin-orbit coupling term with the strength $t_{\text{SO}}$. Here $\hat{c}%
_{j,\sigma }^{\dagger }$ ($\hat{c}_{j,\sigma }$) creates (annihilates) an
atom at site $j$ with spin $\sigma $.

\begin{figure}[t]
\includegraphics[width=3.4in]{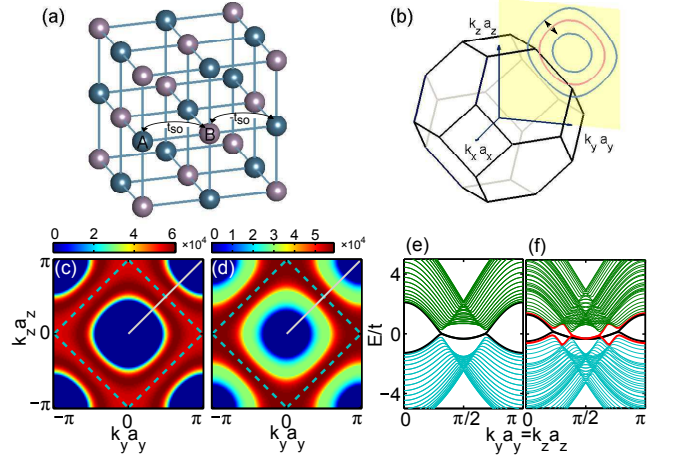} %
\caption{(Color online) {Lattice structure, Brillouin zone, and
surface states.} (a) and (b) Rocksalt lattice structure and corresponding
first Brillouin zone. Dirac ($h_{z}=0$) and Weyl rings ($h_{z}=0.5t$),
denoted by the red and blue rings respectively, are located at the $k_{x}=0$
plane (yellow plane) around ($k_{y}a_{y}=\protect\pi ,k_{z}a_{z}=\protect\pi
$). (c) and (d) Without and with $h_{z}$, density of states at zero energy
without including NNN hopping when there are edges along $x$. The dashed
square indicates the first Brillouin zone. (e) and (f) Spectra along $%
k_{y}a_{y}=k_{z}a_{z}$ [the grey line in (c) and (d)] with edges along $x$
in the presence of NNN hopping, where the black and red lines denote the
surface states. $t_{x}=1.17t$, $t_{y}=t_{z}=t$, $t_{SO}=0.53t$, and $%
t_{N}=-0.07t$. $a_{x}=a_{y}=a_{z}$.}
\label{fig1}
\end{figure}

The position dependent spin-orbit coupling of the Hamiltonian breaks the one
site translation symmetry, leading to a unit cell consisting of two sites: A
and B. These new unit cells form a rocksalt crystal structure as shown in
Fig.~\ref{fig1}(a). In the new basis $\Psi (\mathbf{k})^{T}$ with $\Psi (%
\mathbf{k})=(%
\begin{array}{cccc}
e^{ik_{x}a_{x}}\hat{A}_{\mathbf{k}\uparrow } & e^{ik_{x}a_{x}}\hat{A}_{%
\mathbf{k}\downarrow } & \hat{B}_{\mathbf{k}\uparrow } & \hat{B}_{\mathbf{k}%
\downarrow }%
\end{array}%
)$, the Hamiltonian in the momentum space takes the form
\begin{equation}
H(\mathbf{k})=h_{Nt}-h_{t}\tau _{x}+h_{z}\sigma _{z}+d_{x}\tau _{y}\otimes
\sigma _{y},  \label{HTk}
\end{equation}%
where $h_{Nt}=-2\sum_{\nu }t_{N\nu }\cos (2k_{\nu }a_{\nu })$, $%
h_{t}=2\sum_{\nu }t_{\nu }\cos (k_{\nu }a_{\nu })$ and $d_{x}=2t_{SO}\sin
(k_{x}a_{x})$. $\tau $ are Pauli matrices for the A, B sublattice space. In
the absence of Zeeman fields ($h_{z}=0$), this Hamiltonian preserves both
the pseudo-time-reversal symmetry $\mathcal{T}^{-1}H\mathcal{T}=H(-\mathbf{k}%
)$ with $\mathcal{T}=i\tau _{x}\otimes \sigma _{y}\mathcal{K}$~\cite%
{footnote1} and $\mathcal{K}$ being the complex conjugate operator, and the
inversion symmetry $\mathcal{I}^{-1}H\mathcal{I}=H(-\mathbf{k})$ with $%
\mathcal{I}=\tau _{x}$. These two symmetries guarantee that the state at
each $\mathbf{k}$ is at least two-fold degenerate, which implies that a
gapless touching point, if exists, is four-fold degenerate. Therefore a ring
formed by such gapless points is a Dirac ring. When one of the symmetries is
broken, for instance, $h_{z}$ breaks the pseudo-time-reversal symmetry, a
Dirac ring splits into two Weyl rings, as visualized in Fig.~\ref{fig1}(b).

The emergence of Dirac and Weyl rings can be seen from the energy spectrum
of $H(\mathbf{k})$: $E_{\lambda }=h_{Nt}\pm \sqrt{(h_{z}+\lambda
h_{t})^{2}+d_{x}^{2}}$ with $\lambda =\pm $. When $d_{x}=0$ and $%
h_{z}+\lambda h_{t}=0$, two bands (four bands when $h_{z}=0$) touch to form
nodal rings in the \textbf{k} space. In particular, for $%
-2t_{1}<h_{z}<2t_{2} $ or $-2t_{2}<h_{z}<2t_{1}$ with $%
t_{1}=t_{y}+t_{z}+t_{x}$ and $t_{2}=t_{y}+t_{z}-t_{x}$, such rings emerge in
the $k_{x}=0$ plane as shown in Fig.~\ref{fig1}(b) (the rings in other
planes can be obtained by translating the rings in this plane by a
reciprocal vector). Clearly, when $h_{z}=0$ and $t_{y}+t_{z}>t_{x}$, a
four-fold degenerate Dirac ring appears. With $h_{z}$, the Dirac ring splits
into Weyl rings, whose number equals to the number of the above conditions
satisfied. Around a point on a nodal ring, the energy dispersion is linear
except along the tangent direction to the ring. At the critical points
(i.e., $h_{z}=\pm 2t_{1}$, $\pm 2t_{2}$), a ring shrinks to a point around
which the dispersion is quadratic.

To discuss the topology of these nodal rings, we transform the tight-binding
Hamiltonian (\ref{HTB}) by $(-1)^{j_{x}+j_{y}+j_{z}}\hat{c}_{j,\uparrow
}\rightarrow \hat{c}_{j,\uparrow }$~\cite{Xiongjun2014PRL}, which transforms
Eq. (\ref{HTk}) to
\begin{equation}
H_{1}(\mathbf{k})=h_{Nt}+d_{z}\sigma _{z}-d_{x}\sigma _{x},  \label{THam}
\end{equation}%
with $d_{z}=h_{t}+h_{z}$. The eigenvalues are $E_{\mathbf{k}}^{\pm
}=h_{Nt}\pm \sqrt{d_{x}^{2}+d_{z}^{2}}$, where $\pm $ refer to the helicity,
the eigenvalue of $H_{1}(\mathbf{k})/\sqrt{d_{x}^{2}+d_{z}^{2}}$.

This transformation simplifies the lattice structure to a simple cubic and
hence enlarges the Brillouin zone so that one nodal ring in the $k_{x}=0$
plane is moved to the $k_{x}a_{x}=\pi $ plane. In this transformed model
that possesses the chiral symmetry, i.e., $\sigma _{y}H(\mathbf{k})\sigma
_{y}=-H(\mathbf{k})$ in the absence of NNN hoppings, we see that the Weyl
ring can be characterized by the winding number $n_{w}=1$~\cite%
{Sumata2012PRL}, the number of rotations that the vector $\mathbf{d}=d_{z}%
\mathbf{e}_{x}-d_{x}\mathbf{e}_{y}$ undergoes when it travels along a closed
trajectory enclosing any gap closing point. Such nonzero $n_{w}$ also
amounts to the quantized Berry phase $C_{1}\text{ mod }2\pi =\pi $, half of
the solid angle that $\mathbf{d}$ winds~\cite{Xiao}. For a Dirac ring, the
Hamiltonian (\ref{HTk}) ($h_{z}=0$) respects a $\sigma _{y}$ symmetry, i.e.,
$\sigma _{y}H(\mathbf{k})\sigma _{y}=H(\mathbf{k})$, and hence each band in
two subspaces with different eigenvalues $\sigma _{y}=\pm 1$ has a quantized
Berry phase~\cite{Fan2013PRL,Fan2014PRL}. We note that although the NNN
hopping breaks the chiral symmetry by changing the eigenvalues, it does not
modify the eigenstates, thereby leaving the quantized Berry phase unchanged.

In Fig.~\ref{fig1}(c) and (d), we plot the surface density of states at zero
energy (without NNN hoppings) when the edges are imposed along the $x$
direction in the model~(\ref{HTB}). The density of states is extremely large
between rings in different Brillouin zones, implying the vanishing
dispersion of the surface states (i.e., the surface spectrum is flat). In
the presence of NNN hoppings, the surface spectrum gains a slight dispersion
as shown in Fig.~\ref{fig1}(e) and (f) where the black and red lines denote
the surface spectra. Without $h_{z}$, the surface spectra are four-fold
degenerate, whereas with $h_{z}$, this four-fold degeneracy is lifted so
that the surface states connecting different pairs of gapless points are
separated (black and red lines). This breaking is also reflected in Fig.~\ref%
{fig1}(d) where the density of states in the red region is twice as large as
that in the green one.

\textit{Superfluids in nodal ring lattices}: The Dirac and Weyl nodal ring
lattices can be realized for both Bose and Fermi atoms. Here we consider
fermionic cold atoms with contact attractive interactions that can be tuned
by Feshbach resonances. With attractive interactions, Fermi gases form
superfluids. Under the mean-field approximation, we can define the order
parameter for both A and B sublattices, respectively, as $\Delta
_{A}=-U\langle \hat{A}_{j\downarrow }\hat{A}_{j\uparrow }\rangle $ and $%
\Delta _{B}=-U\langle \hat{B}_{j\downarrow }\hat{B}_{j\uparrow }\rangle $
with the interaction strength $U$ ($U>0$). The dynamics of the superfluid is
governed by the Bogliubov-de Gennes (BdG) Hamiltonian
\begin{equation}
H_{\text{BdG}}=-\tau _{N,z}\otimes (h_{t}\tau _{x}+\tilde{\mu})+d_{x}\tau
_{y}\otimes \sigma _{y}+h_{z}\sigma _{z}+H_{\text{BCS}},
\end{equation}%
where $\tilde{\mu}=\mu -h_{Nt}$ with the chemical potential $\mu $, and $%
\tau _{N}$ act on the Nambu particle-hole space. This Hamiltonian is written
in the Nambu basis $(%
\begin{array}{cc}
\Psi (\mathbf{k}) & \tilde{\Psi}(\mathbf{k})%
\end{array}%
)^{T}$ with $\tilde{\Psi}(\mathbf{k})=(%
\begin{array}{cccc}
e^{ik_{x}a_{x}}\hat{A}_{-\mathbf{k}\downarrow }^{\dagger } & -e^{ik_{x}a_{x}}%
\hat{A}_{-\mathbf{k}\uparrow }^{\dagger } & \hat{B}_{-\mathbf{k}\downarrow
}^{\dagger } & -\hat{B}_{-\mathbf{k}\uparrow }^{\dagger }%
\end{array}%
)$. $\Delta _{A}$ and $\Delta _{B}$ are obtained by numerically solving the
nonlinear gap equations, see Methods.

\begin{figure}[t!]
\includegraphics[width=3.4in]{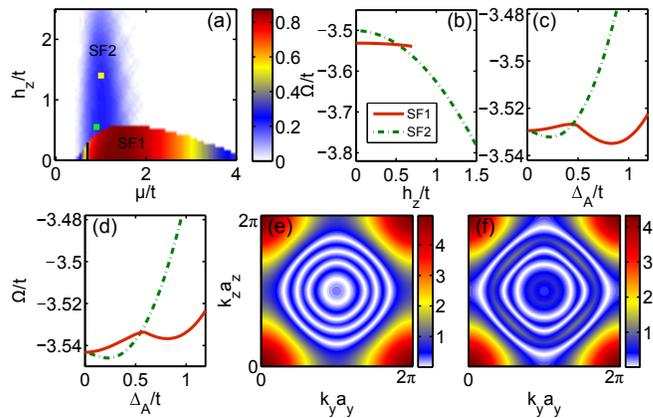} %
\caption{(Color online) {Phase diagram, thermodynamical potential,
and the gap distribution.} (a) The order parameter $\Delta_A$ as a function
of $\protect\mu$ and $h_z$ at zero temperature. SF1 and SF2 correspond to
the phases with $\Delta_A=\Delta_B$ and $\Delta_A=-\Delta_B$, respectively.
A black line divides SF1 phase into gapless (left part) and gapped (right
part) regions, while the whole SF2 phase is gapless. (b) The thermodynamical
potential $\Omega$ of the SF1 (solid red line) and SF2 (dotted-dashed green
line) phases with respect to $h_z$ with $\protect\mu=1.1t$. Note that when $%
h_z>0.7t$ the SF1 state is no longer an energy minimum state, while the SF2
state is in the whole region. (c) and (d) $\Omega$ as a function of $%
\Delta_A $ for the SF1 (solid red line) and SF2 (dotted-dashed line) for $%
\protect\mu=1.1t,h_z=0.5t$ and $\protect\mu=1.1t, h_z=0.6t$, respectively.
(e) and (f) The gap of the quasiparticle spectrum with respect to $k_y a_y$
and $k_z a_z$ in the $k_x=0$ plane for the parameters denoted by the green
and yellow squares in (a), respectively. Here $t_x=1.17t$, $t_y=t_z=t$, $%
t_{SO}=0.53t$, $t_N=-0.07t$, and $U=-4t$. $a_x=a_y=a_z$.}
\label{phase}
\end{figure}

Before we show the numerical results of $\Delta _{A}$ and $\Delta _{B}$, we
first analyze the conditions under which Dirac and Weyl rings can emerge in
the quasiparticle spectrum. Our numerical results show that real $\Delta
_{A} $ and $\Delta _{B}$ with $\Delta _{A}=|\Delta _{B}|$ are energetically
preferred, therefore we only need to consider two superfluids phases: $%
\Delta _{A}=\Delta _{B}$ (dubbed SF1) and $\Delta _{A}=-\Delta _{B}$ (dubbed
SF2), associated with $H_{{\text{B}CS}}=\Delta _{A}\tau _{N,x}$ and $H_{{%
\text{B}CS}}=\Delta _{A}\tau _{N,x}\otimes \tau _{z}$, respectively. When $%
h_{z}=0$, both phases preserve the pseudo-time-reversal and inversion
symmetries (see Methods), which guarantee that the quasiparticle spectra are
at least two-fold degenerate at each $\mathbf{k}$. Therefore gapless rings,
if exist, are four-fold degenerate Dirac rings because of these two
symmetries and the intrinsic particle-hole symmetry. Finite $h_{z}$ breaks
the pseudo-time-reversal symmetry and splits the Dirac ring into two
two-fold degenerate Weyl rings.

Specifically, for the SF1 state, in the absence of $h_{z}$, the eigenvalues
of $H_{\text{BdG}}$ are $E_{\mathbf{k\pm }}^{\lambda }=\pm \sqrt{%
h_{0}^{2}+h_{t}^{2}+d_{x}^{2}+2\lambda \sqrt{\tilde{\mu}%
^{2}h_{t}^{2}+h_{0}^{2}d_{x}^{2}}}$ with $h_{0}^{2}=\Delta _{A}^{2}+\tilde{%
\mu}^{2}$ and $\lambda =\pm $. Each spectrum is two-fold degenerate. From $%
(E_{\mathbf{k}+}^{+})^{2}(E_{\mathbf{k}%
+}^{-})^{2}=(h_{0}^{2}-h_{t}^{2}-d_{x}^{2})^{2}+4\Delta _{A}^{2}h_{t}^{2}$,
we see that $h_{t}=0$ and $d_{x}^{2}=h_{0}^{2}$ for gapless rings. The
latter condition requires $\mu ^{2}\leq 4t_{SO}^{2}-\Delta _{A}^{2}$, if NNN
hoppings are neglected. For the SF2 state, $E_{\mathbf{k}\pm }^{\lambda \nu
}=\pm \sqrt{h_{0}^{2}+(h_{z}+\nu h_{t})^{2}+d_{x}^{2}+2\lambda \sqrt{%
h_{0}^{2}(h_{z}+\nu h_{t})^{2}+\tilde{\mu}^{2}d_{x}^{2}}}$ with $\nu =\pm $.
When $h_{z}=0$, the spectra are two-fold degenerate, and this degeneracy is
explicitly broken by $h_{z}$. Still, by $(E_{\mathbf{k}+}^{+\nu })^{2}(E_{%
\mathbf{k}+}^{-\nu })^{2}=[-h_{0}^{2}+(h_{z}+\nu
h_{t})^{2}+d_{x}^{2}]^{2}+4\Delta _{A}^{2}d_{x}^{2}$, we see that nodal
rings appear when $d_{x}=0$ and $(h_{z}+\nu h_{t})^{2}=h_{0}^{2}$. This
leads to the existence of rings in the $k_{x}=0$ plane when $%
-2t_{1}+h_{0}<h_{z}<2t_{2}+h_{0}$ or $-2t_{1}-h_{0}<h_{z}<2t_{2}-h_{0}$ or $%
-2t_{2}+h_{0}<h_{z}<2t_{1}+h_{0}$ or $-2t_{2}-h_{0}<h_{z}<2t_{1}-h_{0}$, if
NNN hoppings are not involved. For the Weyl rings ($h_{z}\neq 0$), their
number equals to the number of the above relations satisfied. The rings in
other planes are associated with those in the $k_{x}=0$ plane by reciprocal
vectors. The above conditions allow at most two Weyl rings when $%
h_{0}>2(t_{y}+t_{z})$. We note that without $h_{z}$, the rings are Dirac
rings, which split into Weyl rings when $h_{z}$ is turned on. We also note
that NNN hoppings only slightly modify the shape of nodal rings.

In Fig.~\ref{phase}(a), we plot the order parameter $\Delta _{A}$ in the $%
(\mu ,h_{z})$ plane, obtained by solving the nonlinear gap equations at zero
temperature. As we have discussed, there exist two superfluid phases: SF1
and SF2. We can understand these two phases in two limits. In the first
limit, we assume $t_{SO}=0$ and clearly $\Delta _{A}=\Delta _{B}$ as $A$ and
$B$ sublattices can now be connected by a translational transformation. In
the transformed model (\ref{THam}), the momenta of Cooper pairs equal to $%
(\pi /a_{x},\pi /a_{y},\pi /a_{z})$ because of the band inversion~\cite%
{ZhengZhen2015}. We note that in the transformed model, SF1 and SF2 are
associated with the order parameter $\Delta
_{j}=(-1)^{j_{x}+j_{y}+j_{z}}\Delta _{A}$ and $\Delta _{j}=\Delta _{A}$,
respectively. In the second limit, we assume $t_{\nu }=t_{N\nu }=0$, and in
the transformed model, the momenta of Cooper pairs are zero, meaning that $%
\Delta _{B}=-\Delta _{A}$ in the original model. Although these two states
can be simultaneously the energy minimum states as shown in Fig.~\ref{phase}%
(c) and (d), the ground state should be the one with the lower energy.
Therefore with the change of parameters, these two phases can transition
from one to another as shown in Fig.~\ref{phase}(b), where the ground state
changes from SF1 to SF2 with increasing $h_{z}$. Clearly, this phase
transition is the first order.

By examining the quasiparticle spectrum, we find that the SF1 phase is
gapless only in a small region (the left part of the black line), whereas
the SF2 phase is gapless in the whole region. In Fig.~\ref{phase}(e) and
(f), we plot the gap (i.e., $\text{min}(|E_{\mathbf{k}\gamma}|)$) of the
quasiparticle spectrum in the $k_x=0$ plane for the parameters associated
with the green and yellow squares in Fig.~\ref{phase}(a), displaying four
and three Weyl rings, respectively. Similar to the nodal rings in the
single-particle spectrum, the number of rings can be tuned by the Zeeman
field or chemical potential.

Since both SF1 and SF2 phases have the chiral symmetry (i.e., $\mathcal{C}%
^{-1}H_{BdG}\mathcal{C}=-H_{BdG}$ with $\mathcal{C}=\sigma _{x}\tau _{N,y}$%
), similar to the single-particle case, we can associate a winding number $%
n_{w}$ to any 1D closed path enclosing a Weyl ring~\cite{Sumata2012PRL},
\begin{equation}
n_{w}=\frac{1}{2\pi i}\int_{\theta =-\pi }^{\theta =\pi }d\theta \frac{d}{%
d\theta }\log \det A(\theta ),
\end{equation}%
where $A(\theta )=H_{0}+i\Delta _{A}\sigma _{y}$ for the SF1 phase and $%
A(\theta )=H_{0}-i\Delta _{A}\tau _{y}\otimes \sigma _{y}$ for the SF2
phase, with $H_{0}=-(h_{t}\tau _{x}+\tilde{\mu})+d_{x}\tau _{z}\otimes
\sigma _{x}+h_{z}\sigma _{z}$ and $k_{\nu }=k_{\nu }(\theta )$ referring to
a 1D closed path. We find $n_{w}=\pm 1$ for Weyl rings and the associated
Berry phase is $\pm \pi $. In the SF1 phase, when $h_{z}=0$, there exist
Dirac rings, which can be characterized by the winding number in the two
subspaces with different eigenvalues $\sigma _{y}=\pm 1$.

\textit{Realization of nodal ring lattices}: We propose an experimental
setup (shown in Fig.~\ref{fig2}) based on Raman coupling scheme for
generating spin-orbit coupling \cite%
{Lin2011Nature,Jing2012PRL,Zwierlen2012PRL,PanJian2012PRL,Qu2013PRA,Spilman2013PRL, Jing2015arXiv}
to engineer the Hamiltonian~(\ref{WeylH}).

\begin{figure}[t]
\includegraphics[width=3.4in]{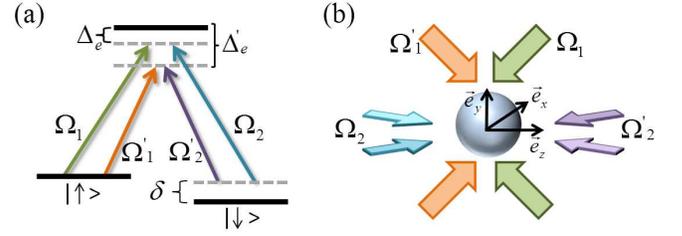}
\caption{(Color online) {Schematics of a laser configuration to
realize the Hamiltonian (\protect\ref{WeylH}).} $\Omega _{1}$ and $\Omega
_{2}$, $\Omega _{1}^{\prime }$ and $\Omega _{2}^{\prime }$ are two sets of
Raman laser beams coupling two hyperfine states $|\uparrow \rangle $ and $%
|\downarrow \rangle $. $\Delta _{e}$ and $\Delta _{e}^{\prime }$ are the
detunings, and $\protect\delta $ is the two-photon detuning. Each Raman
laser beam consists of two plane wave laser beams as shown in (b). These
Raman laser beams also generate optical lattices via the Stark effects.
Additional laser beams are also employed to create optical lattices along $x$
and $z$.}
\label{fig2}
\end{figure}

Two independent pairs of red-detuned Raman lasers are used to couple two
hyperfine states, such as $|\uparrow \rangle =|F=9/2,m_{F}=9/2\rangle $ and $%
|\downarrow \rangle =|F=9/2,m_{F}=7/2\rangle $ for $^{40}$K atoms. One pair
corresponds to the laser beams $\Omega _{1}$ and $\Omega _{2}$ with the Rabi
frequencies $\Omega _{1}=\Omega _{10}\cos (k_{Ry}y)e^{-ik_{Rz}z/2}$ and $%
\Omega _{2}=i\Omega _{20}\sin (k_{Rx}x)e^{ik_{Rz}z/2}$, each of which can be
generated by two plane wave laser beams. The other pair of Raman laser beams
have the Rabi frequencies $\Omega _{1}^{\prime }=\Omega _{10}^{\prime }\cos
(k_{Ry}y)e^{ik_{Rz}z/2}$ and $\Omega _{2}^{\prime }=i\Omega _{20}^{\prime
}\sin (k_{Rx}x)e^{-ik_{Rz}z/2}$, respectively. The detunings $\Delta _{e}\gg
\Omega =|\Omega _{10}\Omega _{20}|/\Delta _{e}$ and $\Delta _{e}^{\prime
}\gg \Omega ^{\prime }=|\Omega _{10}^{\prime }\Omega _{20}^{\prime }|/\Delta
_{e}^{\prime }$ to neglect the population in the excited states by the Raman
procedure. The independence of two Raman coupling pairs are satisfied by $%
|\Delta _{e}-\Delta _{e}^{\prime }|\gg \Omega ,\Omega ^{\prime }$. These
requirements are naturally satisfied in experiments (see the specific
parameters in a typical experiment). Such two sets of Raman laser beams give
rise to the spin-dependent lattice with $\Omega _{SO}=2\Omega $ and $k_{L\nu
}=k_{R\nu }$ in Eq.~(\ref{WeylH}) when $\Omega =\Omega ^{\prime }$, achieved
when the two sets of Raman lasers come from the same resource. These laser
beams also lead to optical lattices along the $x$ and $y$ directions via the
stark effects: $-\delta V_{x}\sin ^{2}(k_{Rx}x)$ and $-V_{y}\cos
^{2}(k_{Ry}y)$ with $\delta V_{x}=(|\Omega _{20}|^{2}+|\Omega _{20}^{\prime
}|^{2})/\Delta _{e}$ and $V_{y}=(|\Omega _{10}|^{2}+|\Omega _{10}^{\prime
}|^{2})/\Delta _{e}$ (we have assumed $\Delta _{e}\approx \Delta
_{e}^{\prime }>0$ given that $\Delta _{e}$ and $\Delta _{e}^{\prime }$ are
both in the order of THz whereas $|\Delta _{e}-\Delta _{e}|^{\prime }$ in
the order of 10-100MHz). Moreover, one needs another stronger optical
lattices along the $x$ direction: $-V_{x}^{\prime }\cos ^{2}(k_{Rx}x)$ with $%
V_{x}^{\prime }>0$ so that the total $x$ direction optical lattice is $%
-V_{x}\cos ^{2}(k_{Rx}x)$ with $V_{x}=V_{x}^{\prime }-\delta V_{x}>0$.
Similarly, optical lattices along $z$, $-V_{z}\cos ^{2}(k_{Rz}z)$ with $%
V_{z}>0$, can be generated. We note that the Raman laser beams can also
create the Zeeman field $h_{z}=\delta /2$ with $\delta $ being the
two-photon detuning.

In experiments, we consider $^{40}$K atoms and choose $\Delta _{e}=2\pi
\times 1.46$THz that can be realized by a red-detuned laser beam with
wavelength 773 nm~\cite{Jing2012PRL}, which gives the recoil energy $%
E_{R}/\hbar =2\pi \times 8.3$kHz. A simple geometry of laser beams gives
rise to $k_{Rx}=k_{Ry}=k_{Rz}=\sqrt{4/5}k_{R}$. The two pairs of Raman laser
beams are independent as $|\Delta _{e}-\Delta _{e}^{\prime }|\sim 2\pi
\times (10-100)$MHz $\gg E_{R}$. For $\Omega _{10}=\Omega _{10}^{\prime
}=2\pi \times 0.14$ GHz and $\Omega _{20}=\Omega _{20}^{\prime }=2\pi \times
0.035$ GHz, we have $\Omega _{SO}=0.8E_{R}$, $V_{y}=3.2E_{R}$, and $\delta
V_{x}=V_{y}/16$. For $\Omega _{3x}=\Omega _{3z}=2\pi \times 0.21$ GHz, we
have $V_{x}^{\prime }=V_{z}=V_{y}$. $\delta $ can be readily tuned from
zero. With these parameters, in the tight-binding model, $t=0.068E_{R}$, $%
t_{x}=1.17t$, $t_{y}=t_{z}=t$, $t_{N\nu }=t_{N}=-0.07t$, and $t_{SO}=0.53t$.
The Dirac and Weyl rings in the single particle spectrum can be detected
through the spin-resolved radio-frequency spectroscopy, similar to that in
spin-orbit coupled atomic gases \cite%
{Jing2012PRL,Zwierlen2012PRL,Jing2015arXiv}. The rings in the superfluids
may be detected by measuring the spectral density~\cite{Melo2011PRL,Yong2015}
using the momentum resolved photoemission spectroscopy \cite{Jin2008}. In
terms of a BEC loaded in nodal ring lattices, one can measure Landau-Zener
tunneling probability to detect the rings~\cite{Esslinger2012Nature,Law2015}
and the interference between two BECs traveling across a Weyl ring to
extract the Berry phase~\cite{Bloch2014Science}.


\textit{Discussion}: Not only nodal rings in 3D can be realized in the proposed experimental
setup, but also Dirac cones in 2D can be engineered in a much simpler setup
with only a pair of Raman laser beams (see supplementary information). In
contrast to 2D Dirac cones in honeycomb lattices in previous experiments \cite%
{Esslinger2012Nature}, there are two types of Dirac cones: one with
four-fold degeneracies (each with Berry phase being $\pi $ or $-\pi $ in the
subspaces with $\sigma _{y}=\pm 1$ similar to the 3D case) and one with
two-fold degeneracies (each with Berry phase being $\pi $ or $-\pi $).
Without Zeeman fields, the former can exist, while with Zeeman fields, the
former splits into two Dirac cones with two-fold degeneracies in separated
positions in the momentum space. Such Dirac cones can be readily created,
moved, and merged by tuning the lattice strength and Zeeman fields. Note
that in previous experiments \cite{Esslinger2012Nature}, the Dirac cones are
formed due to the honeycomb lattice structure and the spin Zeeman field only
shifts the relative energy between two Dirac cones for different spins, not
their positions in the momentum space. In our model, the Zeeman field can
split a four-fold degenerate Dirac cone into two located at different
positions in the momentum space (see supplementary information).

In summary, we propose an experimental setup to engineer an optical lattice
system that support nodal rings (i.e., Dirac or Weyl rings) in its
single-particle spectrum. We study the superfluidity of Fermi gases with
attractive interactions in such a lattice and show that the quasiparticle
spectrum can also exhibit the nodal rings. Our scheme is based on previously
already successful experimental setup and should pave the way for the
experimental generation and observation of topological gapless materials.


\begin{center}
Methods
\end{center}

\textit{BdG equation in momentum space}: Here we only consider the BCS
pairing with zero center-of-mass momenta Cooper pairs and hence $\Delta _{A}$
and $\Delta _{B}$ are spatially uniform. With a global $U(1)$ gauge
invariance, although we can choose positive $\Delta _{A}$ and complex $%
\Delta _{B}$ for a general case, our numerical calculation shows that being
real of $\Delta _{B}$ is energetically preferred. Therefore, with real $%
\Delta _{B}$, $H_{\text{BCS}}=\tau _{N,x}\otimes (\Delta _{A}\tau
_{z}^{+}+\Delta _{B}\tau _{z}^{-})$ with $\tau _{z}^{\pm }=(\tau _{0}\pm
\tau _{z})/2$. The thermodynamical potential per site at the temperature $%
T=1/(k_{B}\beta )$ with Boltzmann constant $k_{B}$ is
\begin{equation}
\Omega =\frac{1}{U}|\Delta _{0}|^{2}-\sum_{\mathbf{k}}\left[ 2\mu +\frac{1}{%
\beta }\sum_{\gamma }\frac{1}{2}\ln (1+e^{-\lambda \beta E_{\mathbf{k}\gamma
}})\right] ,
\end{equation}%
with $\Delta _{0}^{2}=\Delta _{A}^{2}+\Delta _{B}^{2}$ and $E_{\mathbf{k}%
\gamma }$ with $\gamma =1,2,\cdots ,8$ being the eigenvalues of $H_{\text{BdG%
}}$. To obtain the mean-field order parameters, we solve the nonlinear gap
equations
\begin{equation}
\begin{array}{cc}
\partial \Omega /\partial \Delta _{A}=0, & \partial \Omega /\partial \Delta
_{B}=0.%
\end{array}%
\end{equation}%
For SF1 and SF2 phases, the pseudo-time-reversal symmetries correspond to $%
\mathcal{T}_{1}^{-1}H_{\text{BdG}}\mathcal{T}_{1}=H_{\text{BdG}}(-\mathbf{k}%
) $ and $\mathcal{T}_{2}^{-1}H_{\text{BdG}}\mathcal{T}_{2}=H_{\text{BdG}}(-%
\mathbf{k})$ with $\mathcal{T}_{1}=\tau _{N,0}\otimes \mathcal{T}$ ($\tau
_{N,0}$ is a 2$\times $2 identity matrix) and $\mathcal{T}_{2}=\tau
_{N,z}\otimes \mathcal{T}$, respectively. The inversion symmetries
correspond to $\mathcal{I}_{1}^{-1}H_{\text{BdG}}\mathcal{I}_{1}=H_{\text{BdG%
}}(-\mathbf{k})$ and $\mathcal{I}_{2}^{-1}H_{\text{BdG}}\mathcal{I}_{2}=H_{%
\text{BdG}}(-\mathbf{k})$ with $\mathcal{I}_{1}=\tau _{N,0}\otimes \mathcal{I%
}$ and $\mathcal{I}_{2}=\tau _{N,z}\otimes \mathcal{I}$, respectively. The
particle-hole symmetry is associated with the transformation ${\Xi }%
_{j}^{-1}H{\Xi }_{j}=-H(-\mathbf{k})$ ($j=1,2$) with ${\Xi }_{1}=i\tau
_{N,y}\otimes \tau _{x}\otimes \sigma _{y}\mathcal{K}$ and ${\Xi }_{2}=i\tau
_{N,x}\otimes \tau _{x}\otimes \sigma _{y}\mathcal{K}$ for SF1 and SF2,
respectively.

\textbf{Acknowledgements}: We would like to thank P. Engels, F. Zhang, and
C. Liu for helpful discussions. Y.X. and C.Z are supported by ARO
(W911NF-12-1-0334), AFOSR (FA9550-13-1-0045), and NSF (PHY-1505496). We also
thank Texas Advanced Computing Center, where part of our numerical
calculations was performed.

\textbf{Author contributions} All authors took part in the discussing of the
results, designing of the experiment setup, and the writing of the
manuscript. Y. Xu conceived the idea and obtained the numerical results. C.
Zhang supervised the project.

\textbf{Competing financial interests}

The authors declare no competing financial interests.

\begin{widetext}
\maketitle

\setcounter{equation}{0} \setcounter{figure}{0} \setcounter{table}{0} %
\renewcommand{\theequation}{S\arabic{equation}} \renewcommand{\thefigure}{S%
\arabic{figure}} \renewcommand{\bibnumfmt}[1]{[S#1]} \renewcommand{%
\citenumfont}[1]{S#1}

\section{S-1. DERIVATION OF TIGHT-BINDING MODEL}

In this supplementary material, we derive the tight-binding model from the
continuous model (1) in the main text and compare the single-particle
spectra of the tight-binding and continuous models for typical parameters in
experiments.

\begin{figure}[t]
\includegraphics[width=6in]{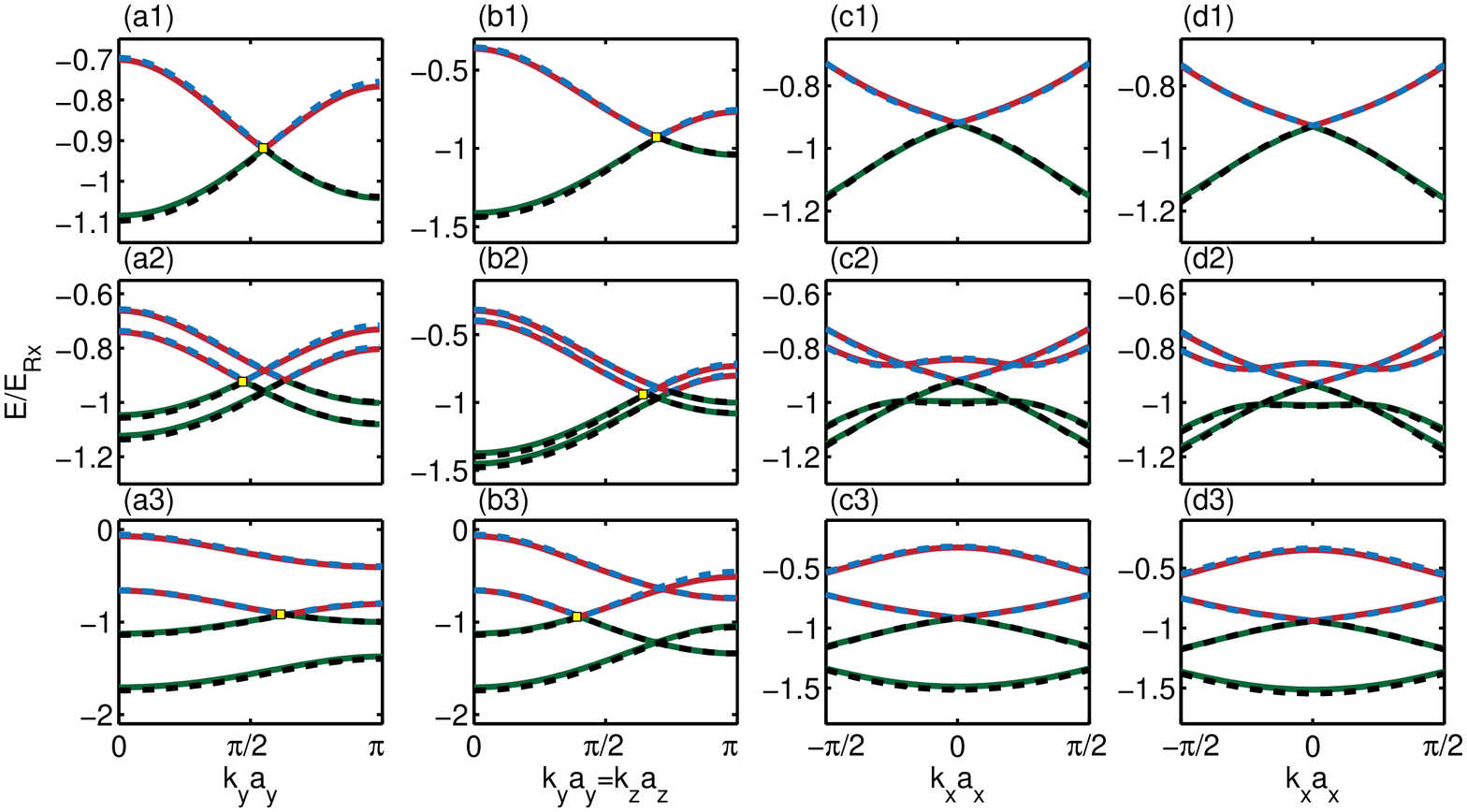}
\caption{(Color online) Single-particle spectra obtained by the
tight-binding model (dashed blue and black lines) and the continuous model
(solid red and green lines). Each row panel from top to bottom is associated
with $h_z=0$, $h_z=0.032E_{R}$, and $h_z=0.24E_{R}$, respectively. The first
and second column panels (from left to right) correspond to $(k_x=0,k_z a_z=%
\protect\pi)$ and $(k_x=0)$, respectively. The third and fourth column
panels plot the spectra along $k_x$ around touching points (yellow squares)
corresponding to the first and second column panels, respectively. The
parameters for the continuous model are $k_{Lx}=k_{Ly}=k_{Lz}=\protect\sqrt{%
4/5}k_R$, $V_x=V_y=V_z=3.2E_R$, and $\Omega_{SO}=0.8E_R$; the parameters for
the tight-binding model are $t=t_y=t_z=0.068E_R$ and $t_N=-0.07t$ and $%
t_{SO}=0.53t$. The recoil energy along $x$ is $E_{Rx}=\hbar^2
k_{Lx}^2/2m=0.8E_R$. }
\label{SIfig1}
\end{figure}

In the second quantization representation, the Hamiltonian takes the form
\begin{equation}
H_{II}=\int d\mathbf{r}\hat{\psi}^{\dagger }(\mathbf{r})H\hat{\psi}(\mathbf{r%
}),
\end{equation}%
where $H$ is the single-particle Hamiltonian in Eq.(1) in the main text, $%
\hat{\psi}(\mathbf{r})=(%
\begin{array}{cc}
\hat{\psi}_{\uparrow }(\mathbf{r}) & \hat{\psi}_{\downarrow }(\mathbf{r})%
\end{array}%
)^{T}$ where $\hat{\psi}_{\sigma }(\mathbf{r})$ [$\hat{\psi}_{\sigma
}^{\dagger }(\mathbf{r})$] annihilates (creates) an atom with spin $\sigma $
($\sigma =\uparrow ,\downarrow $) located at $\mathbf{r}$. They satisfy the
anti-commutation or commutation relation $[\hat{\psi}_{\sigma }(\mathbf{r}),%
\hat{\psi}_{\sigma ^{\prime }}^{\dagger }(\mathbf{r}^{\prime })]_{\pm
}=\delta _{\sigma \sigma ^{\prime }}\delta (\mathbf{r}-\mathbf{r}^{\prime })$
for fermionic atoms ($+$) or bosonic atoms ($-$), respectively. The field
operator can be expanded by local Wannier functions
\begin{equation}
\hat{\psi}_{\sigma }(\mathbf{r})=\sum_{nj\sigma }W_{nj\sigma }\hat{c}%
_{n,j,\sigma },
\end{equation}%
where $W_{nj\sigma }$ is the Wannier function located at the $j$-th site for
the $n$-th band for spin $\sigma $, and $\hat{c}_{n,j,\sigma }$ annihilates
an atom at the $j$-th site in the $n$-th band with spin $\sigma $. As we
only consider the physics in the lowest band, let us assume $n=1$ and
further assume that the Wannier function $W_{1j\sigma }$ can be approximated
by the lowest band Wannier function $W_{j}$ of the Hamiltonian with pure
spin-independent optical lattices. Hence
\begin{equation}
\hat{\psi}_{\sigma }(\mathbf{r})\approx \sum_{j}W_{j}\hat{c}_{j,\sigma },
\end{equation}%
where $W_{j}=W_{j_{x}}^{x}(r_{x})W_{j_{y}}^{y}(r_{y})W_{j_{z}}^{z}(r_{z})$
with $W_{j_{\nu }}^{\nu }(r_{\nu })=W^{\nu }(r_{\nu }-j_{\nu }a_{\nu })$
being the Wannier function along $\nu $. Based on this expansion, the
tight-binding model without $H_{SO}$ reads
\begin{equation}
H_{1}=-\sum_{j}\sum_{\sigma }\sum_{\nu }(t_{\nu }\hat{c}_{j,\sigma
}^{\dagger }\hat{c}_{j_{\nu }+1,\sigma }+t_{N\nu }\hat{c}_{j,\sigma
}^{\dagger }\hat{c}_{j_{\nu }+2,\sigma }+h.c.)+h_{z}\sum_{j}(\hat{c}%
_{j,\uparrow }^{\dagger }\hat{c}_{j,\uparrow }-\hat{c}_{j,\downarrow
}^{\dagger }\hat{c}_{j,\downarrow })
\end{equation}%
with the inclusion of the nearest and next nearest neighbor hopping with the
corresponding hopping amplitudes being
\begin{eqnarray}
t_{\nu } &=&-\int dr_{\nu }W_{j}\left[ \frac{p_{\nu }^{2}}{2m}-V_{\nu }\cos
^{2}(k_{L\nu }r_{\nu })\right] W_{j_{\nu }+1},  \label{ET} \\
t_{N\nu } &=&-\int dr_{\nu }W_{j}\left[ \frac{p_{\nu }^{2}}{2m}-V_{\nu }\cos
^{2}(k_{L\nu }r_{\nu })\right] W_{j_{\nu }+2}.  \label{ETN}
\end{eqnarray}%
The tight-binding term contributed by the spin-dependent lattices can be
derived as follows
\begin{eqnarray}
H_{SO} &=&i\Omega _{SO}\int d\mathbf{r}\hat{\psi}_{\uparrow }^{\dagger }(%
\mathbf{r})V_{SO}\hat{\psi}_{\downarrow }(\mathbf{r})+h.c. \\
&\approx &i\Omega _{SO}\sum_{j,j^{\prime }}\hat{c}_{j,\uparrow }^{\dagger }%
\hat{c}_{j^{\prime },\downarrow }t_{SO}^{jj^{\prime }}+h.c.,
\end{eqnarray}%
where
\begin{equation*}
t_{SO}^{jj^{\prime }}=\int d\mathbf{r}W_{j}V_{SO}W_{j^{\prime }}=\prod_{\nu
=x,y,z}t_{SO}^{j_{\nu }j_{\nu }^{\prime }},
\end{equation*}%
with
\begin{eqnarray}
t_{SO}^{j_{x}j_{x}^{\prime }} &=&t_{SO}^{j_{x}^{\prime }j_{x}}=\int
dr_{x}W_{j_{x}}^{x}(r_{x})\sin (k_{Lx}r_{x})W_{j_{x}^{\prime }}^{x}(r_{x}),
\\
t_{SO}^{j_{y}j_{y}^{\prime }} &=&t_{SO}^{j_{y}^{\prime }j_{y}}=\int
dr_{y}W_{j_{y}}^{y}(r_{y})\cos (k_{Ly}r_{y})W_{j_{y}^{\prime }}^{y}(r_{y}),
\\
t_{SO}^{j_{z}j_{z}^{\prime }} &=&t_{SO}^{j_{z}^{\prime }j_{z}}=\int
dr_{z}W_{j_{z}}^{z}(r_{z})\cos (k_{Lz}r_{z})W_{j_{z}^{\prime }}^{z}(r_{z}).
\end{eqnarray}

Because one of the optical wells is located at $\mathbf{r}=(0,0,0)$, $%
W_{0}^{\nu }(r_{\nu })=W_{0}^{\nu }(-r_{\nu })$ and%
\begin{eqnarray}
&&t_{SO}^{j_{x}j_{x}}=t_{SO}^{j_{x}j_{x}+2}=t_{SO}^{j_{y}j_{y}+1}=t_{SO}^{j_{z}j_{z}+1}=0,
\\
&&t_{SO}^{j_{\nu }j_{\nu }+1}=-t_{SO}^{j_{\nu }+1j_{\nu }+2}, \\
&&t_{SO}^{j_{\nu }j_{\nu }}=-t_{SO}^{j_{\nu }+1j_{\nu }+1},
\end{eqnarray}%
where the last two relations are obtained because the period of the
spin-independent optical lattices is a half of that of the spin-dependent
ones along each direction. Therefore, with the nearest-neighbor hopping (no
next nearest-neighbor hopping exists), the position dependent spin-orbit
coupling term of the tight-binding model reads
\begin{eqnarray}
H_{SO} &=&i\Omega _{SO}\sum_{j}\left[ \hat{c}_{j,\uparrow }^{\dagger }\hat{c}%
_{j_{x}+1,\downarrow
}t_{SO}^{j_{x}j_{x}+1}t_{SO}^{j_{y}j_{y}}t_{SO}^{j_{z}j_{z}}+\hat{c}%
_{j,\uparrow }^{\dagger }\hat{c}_{j_{x}-1,\downarrow
}t_{SO}^{j_{x}j_{x}-1}t_{SO}^{j_{y}j_{y}}t_{SO}^{j_{z}j_{z}}\right] +h.c. \\
&=&i\Omega _{SO}\sum_{j}\left[ \hat{c}_{j,\uparrow }^{\dagger }\hat{c}%
_{j_{x}+1,\downarrow }-\hat{c}_{j,\uparrow }^{\dagger }\hat{c}%
_{j_{x}-1,\downarrow }\right]
t_{SO}^{j_{x}j_{x}+1}t_{SO}^{j_{y}j_{y}}t_{SO}^{j_{z}j_{z}}+h.c. \\
&=&it_{SO}\sum_{j}(-1)^{j_{x}+j_{y}+j_{z}}\left[ \hat{c}_{j,\uparrow
}^{\dagger }\hat{c}_{j_{x}+1,\downarrow }-\hat{c}_{j,\uparrow }^{\dagger }%
\hat{c}_{j_{x}-1,\downarrow }\right] +h.c.
\end{eqnarray}%
where
\begin{equation}
t_{SO}=\Omega _{SO}t_{SO}^{01}t_{SO}^{00}t_{SO}^{00}.  \label{ETSO}
\end{equation}%
Therefore, we obtain the tight-binding model in Eq.(2) in the main text ($%
H_{1}=H_{h}+H_{Z}$).

\begin{figure}[t]
\includegraphics[width=3.4in]{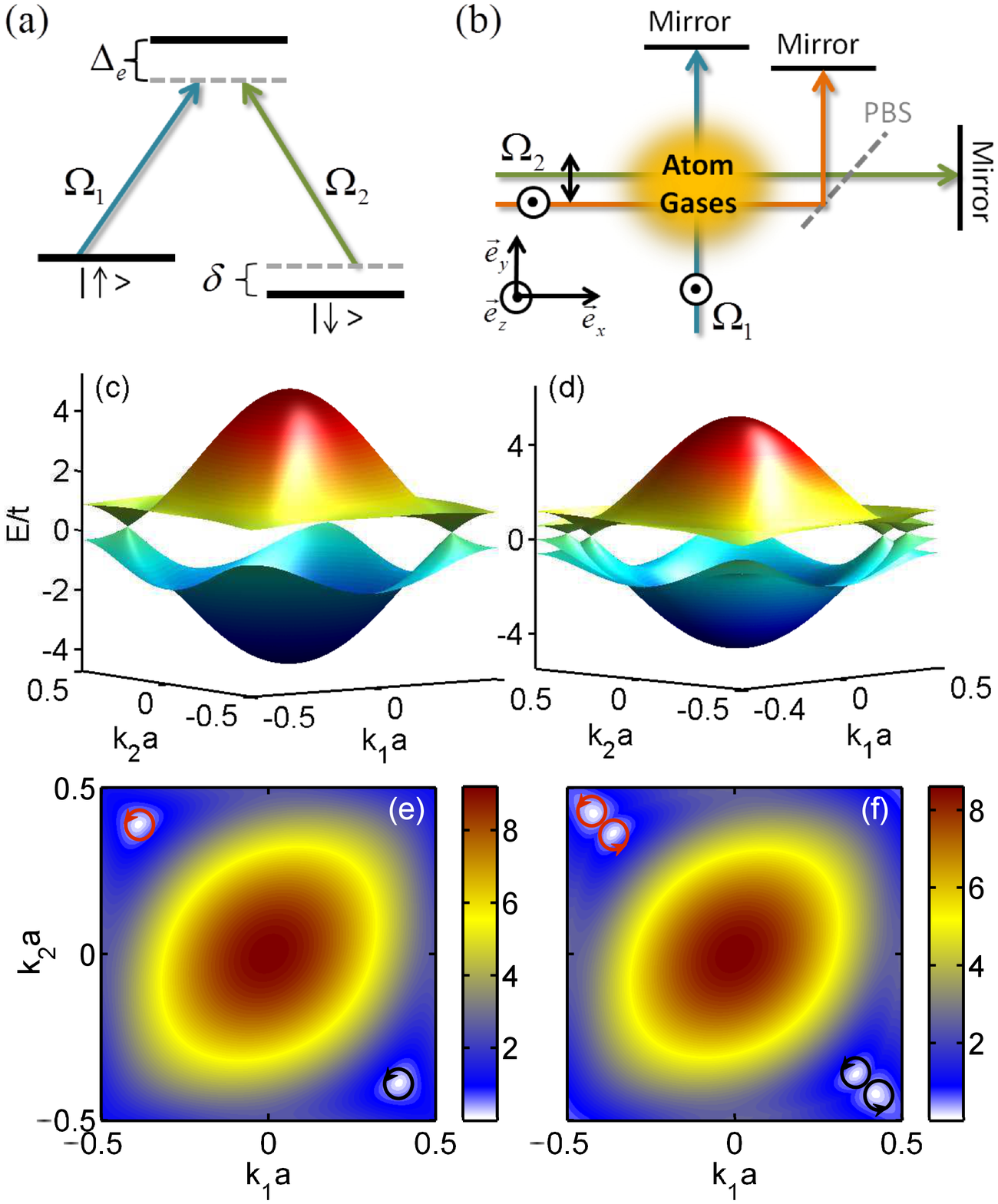}
\caption{(Color online) {Schematics of a laser configuration to
realize the 2D Hamiltonian with Dirac cones and the single-particle spectra
of such optical lattice systems.} $\Omega _{1}$ and $\Omega _{2}$ are two
Raman laser beams coupling two hyperfine states $|\uparrow \rangle $ and $%
|\downarrow \rangle $. $\Delta _{e}$ is the detuning, and $\protect\delta $
is the two-photon detuning. Each Raman laser beam is a standing wave formed
by a plane wave laser beam reflected by a mirror as shown in (b). These
Raman laser beams also generate optical lattices via the Stark effects. An
additional laser beam (red line) with different frequency from the Raman
lasers (shifted by $\sim $100MHz using an acoustic-optical modulator (AOM))
is also employed to create optical lattices along $x$. PBS denotes
polarizing beamsplitter. Double arrow and circle dots denote the
polarization direction of laser beams. The PBS separates the Raman beam and
the optical lattice beams so that their phases can be controlled
individually by different mirrors. (c)(d) Single-particle spectra of the
tight-binding Hamiltonian without and with Zeeman fields, respectively.
(e)(f) The gap distribution between particle and hole branches in the
momentum space. The white points indicate the Dirac cones and the Berry
phase calculated along the red circle (black one) is $\protect\pi $ ($-%
\protect\pi $). Note in (e) the Berry phase is calculated in the subspace of
$\protect\sigma _{y}$. Here $t_{x}=t$, $t_{y}=1.3t$, $t_{N}=0.07t$, and $%
t_{SO}=0.64t$. The lattice constants are $a_{x}=a_{y}=a$ and the crystal
momenta are $a\mathbf{k}=\protect\pi (k_{1}+k_{2})\mathbf{e}_{x}+\protect\pi %
(-k_{1}+k_{2})\mathbf{e}_{y}$. }
\label{SIfig2}
\end{figure}

We now consider a specific set of experimental parameters used in the main
text, yielding $k_{Lx}=k_{Ly}=k_{Lz}=\sqrt{4/5}k_R$, $V_x=V_y=V_z=3.2E_R$,
and $\Omega_{SO}=0.8E_R$. The tight-binding parameters are calculated from
Eq.~(\ref{ET}), Eq.~(\ref{ETN}), and Eq.~(\ref{ETSO}), yielding $%
t=t_y=t_z=0.068E_R$, $t_N=-0.07t$, and $t_{SO}=0.53t$. We note that we
choose $t_x=1.17t$, which is slightly different from $t_y$ and $t_z$ given
the distinct correction of Wannier functions by $V_{SO}$ along $x$. In Fig.~%
\ref{SIfig1}, we compare the single-particle spectra obtained by the
tight-binding model with that obtained by the continuous model, showing
their good agreement with each other.

\section{S-2. DIRAC CONES IN TWO DIMENSIONS}

We consider a two-dimensional case with the following Hamiltonian
\begin{equation}
H_{2D}=\frac{\mathbf{p}^{2}}{2m}-\sum_{\nu =x,y}V_{\nu }\cos
^{2}(k_{R}r_{\nu })+h_{z}\sigma _{z}-V_{SO}\sigma _{y},
\end{equation}%
where $V_{SO}=\Omega _{SO}\sin (k_{L_{R}}r_{x})\cos (k_{L_{R}}r_{y})$.
Compared with the experimental setup for the 3D scenario in Fig. 3 in the
main text, the setup to realize this Hamiltonian is much easier. One only
needs a pair of Raman laser beams [shown in Fig.~\ref{SIfig2} (a) and (b)]
with Rabi frequencies $\Omega _{1}=\Omega _{10}\cos (k_{R}y)$ and $\Omega
_{2}=i\Omega _{20}\sin (k_{R}x)$, and an additional standing laser beam to
engineer an optical lattice along $x$. The tight-binding model of this
Hamiltonian is a simplified version of Eq. (3) when the hopping terms along $%
x$ and $y$ are kept and
\begin{equation}
H_{SO}=it_{\text{SO}}\sum_{j}(-1)^{j_{x}+j_{y}}(\hat{c}_{j,\uparrow
}^{\dagger }\hat{c}_{j_{x}+1,\downarrow }-\hat{c}_{j,\uparrow }^{\dagger }%
\hat{c}_{j_{x}-1,\downarrow })+h.c..
\end{equation}%
Clearly, the Hamiltonian in the momentum space is also a simplified version
of the Hamiltonian (3) in the main text when only the hopping terms along $x$
and $y$ are kept. Dirac cones appear when $h_{z}+\lambda h_{t}=0$ on the $%
k_{x}=0$ line. Different from the 3D case, when $h_{z}=0$, two Dirac cones
[as displayed in Fig.~\ref{SIfig2}(c)] with four-fold degeneracies (each
with Berry phase being $\pi $ or $-\pi $ in the subspaces with $\sigma
_{y}=\pm 1$ similar to the 3D case) can appear only when $t_{y}>t_{x}$,
which can be realized by choosing a stronger optical lattice along the $x$
direction than that along the $y$ direction. At the critical point $%
t_{y}=t_{x}$, the spectrum becomes quadratic along $y$ and keeps linear
along $x$. In the presence of $h_{z}$, each Dirac cone with four-fold
degeneracies splits into two Dirac cones with two-fold degeneracies (each
with Berry phase being $\pi $ or $-\pi $) at different positions in the
momentum space as shown in Fig.~\ref{SIfig2}(d). They can exist as long as $%
2(t_{x}-t_{y})<h_{z}<2(t_{y}+t_{x})$ and $-(t_{x}+t_{y})<h_{z}<2(t_{y}-t_{x})
$. When $h_{z}=\pm 2(t_{x}+t_{y})$ and $h_{z}=\pm 2(t_{y}-t_{x})$, two Dirac
cones with two-fold degeneracies merge---with spectra being quadratic along $%
y$ and linear along $x$---at $(k_{x}=0,k_{y}=0)$ and $(k_{x}=0,k_{y}=\pi )$,
respectively.
\end{widetext}
\end{document}